\documentclass[sigconf]{acmart}

\newif\iftrack
\trackfalse 

\AtBeginDocument{%
  }

\setcopyright{CC}
\copyrightyear{2026}
\acmYear{2026}
\acmDOI{}
\acmConference[GI 26]{Graphics Interface}{June 09--12,
  2026}{Kitchener-Waterloo, Ontario, Canada}

\usepackage{xargs} 
\usepackage{soul}  
\usepackage[nameinlink,capitalise]{cleveref}

\crefname{table}{Table}{Tables}
\crefname{figure}{Figure}{Figures}
\crefname{section}{Section}{Sections}

\usepackage{color} 

\newcommand{\bstart}[1]{\vspace{1mm} \noindent{\textbf{#1.}}}

\newcommand{\istart}[1]{\vspace{1mm} \noindent{\textit{#1.}}}

\newcommand{\ie}{i.e.,\ }
\newcommand{\etal}{et al.}
\newcommand{\eg}{e.g.,\ }

\newcommandx{\revision}[2][1=]{\iftrack{\textcolor{gray}{\st{#1}}\textcolor{red}{#2}}\else{#2}\fi}
\begin{document}

\title[The Choreography of Augmented Reality Timelines]{The Choreography of Augmented Reality Timelines: Studying the Relative Position, Chronology, \& Situatedness of Event Sequences}


\author{Isabelle Kwan}
\orcid{0009-0006-5074-7913}
\affiliation{%
  \institution{Simon Fraser University}
  \city{Burnaby}
  \state{British Columbia}
  \country{Canada}
}
\email{isabelle_kwan@sfu.ca}

\author{Jessica Ziyu Chen}
\orcid{0009-0006-1098-6215}
\affiliation{%
  \institution{University of Waterloo}
  \city{Waterloo}
  \state{Ontario}
  \country{Canada}}
\email{jz3chen@uwaterloo.ca}

\author{Matthew Brehmer}
\orcid{0000-0001-5524-2291}
\affiliation{%
  \institution{University of Waterloo}
  \city{Waterloo}
  \state{Ontario}
  \country{Canada}
}
\email{mbrehmer@uwaterloo.ca}


\begin{abstract}
Timelines are effective ways to tell historical\revision[, biographical,] and personal stories. However, most timeline visualization tools impose an inflexible model of time prioritizing chronological clarity. On the other hand, unconstrained representations can better capture the irregular and contextual nature of lived time, but often at the cost of interpretability. In this work, we explore this continuum with a study of how historical and personal timelines could manifest in physical spaces. We conducted a formative study (N=12) in which participants freely arranged events within a physical environment. We observed a diversity of strategies reflecting the personal and context-dependent nature of temporal mental models. We also invited participants to consider how others could move through their timelines. \revision[suggesting]{Our analysis led to} a choreographic approach to timeline creation\revision[. Based on these insights, we developed]{, as well as} a proof-of-concept tablet-based augmented reality (AR) application that supports spatial timeline drawing and viewing. Finally, we reflect on the design implications of encoding chronology, pacing, and spatial context in immersive timeline stories.
\end{abstract}

\begin{teaserfigure}
  \includegraphics[width=\textwidth]{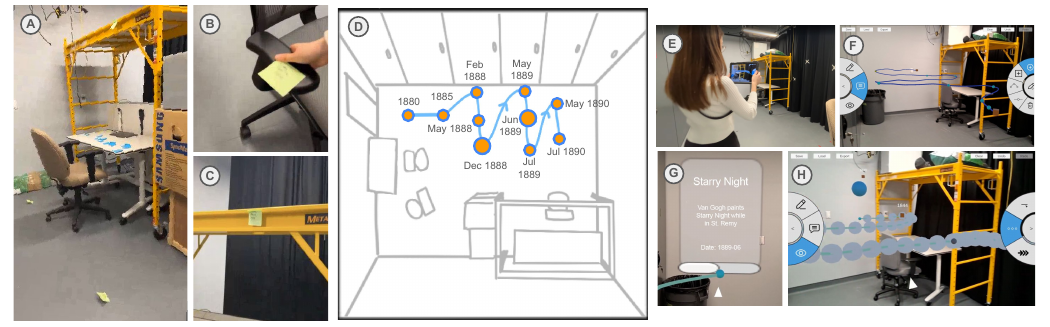}
  \caption{(A---C) In our formative study, participants placed events (dates + descriptions) as sticky notes around a room, which we analyzed via sketching (\eg~(D)). Our analysis informed the design of a (E) tablet-based prototype for (F) the choreography and (G) playback of augmented reality timelines featuring (H) proxemic reveal and animated guidance.}
  \Description{(A) a room containing a chair and a scaffold; (B) a sticky note on the arm of a chair; (C) a sticky note on the scaffold; (D) an overhead sketch of a timeline in the room shown in A-C; (E) a person holds a tablet with both hands, an augmented reality application is on the screen, displaying pass-through video of the room containing the chair and scaffold; (F) a screenshot from the application, showing two radial menus of widgets on either side, and pass-through video of the room augmented with a blue curved spline extending into the room; (G) a virtual information card appears above the a waypoint on the aforementioned spline, and a compass appears at the bottom of the display; (H) particles appear along the spline.}
  \label{fig:teaser}
\end{teaserfigure}


\maketitle

\section{Introduction}
Timelines are widely used to represent sequences of events and communicate temporal structure. 
Most contemporary timeline representations follow a single linear axis~\cite{Rosenberg2010}, mapping events onto fixed intervals of chronological time. 
While this structure supports clarity and comparison, it imposes a strict regularity on temporal representation. 
Historical and especially biographical narratives are often ineffectively conveyed by this form of representation, particularly as stories depicting the lived experiences of people are nonlinear and unevenly distributed in time, calling for a larger palette of timeline design choices~\cite{brehmer2016timelines}.
However, completely unconstrained and idiosyncratic representations of time~\cite{offenwanger2023timesplines,torna2011} can introduce other challenges. 
When people are free to place events without temporal structure, chronological ordering and duration relationships can become difficult to interpret.
This tension brings us to our key design question: how can historical and biographical timeline representations remain interpretable while allowing people to express more flexible and meaningful spatial structures?

The emergence of spatial computing motivates a new approach to answering this question. 
Augmented reality (AR) environments allow information to be placed directly within physical space, enabling timelines that extend beyond the boundaries of a flat display. 
For example, recent work has demonstrated the creation of immersive branching and circular timelines in AR \cite{vu2025walking}.
In this work, we ask a more fundamental and formative question regarding the arrangement of timeline structure in space and the role of spatial familiarity, allowing us to reflect on the implications for creating, sharing, and navigating spatial timeline stories.
Specifically, we probe how people might leverage (familiar) physical surroundings and the objects contained therein when arranging temporal information, as well as the role of chronology in these representations. 

In this paper, we conceptualize the process of creating and sharing timelines in physical space as a form of \textbf{choreography}, whereupon people do not merely arrange events, but rather guide and pace movement through them. 
In this framing, timeline creators implicitly direct where a viewer should go, how long they might linger at a point in space to appreciate the events placed there, and how transitions between events are experienced, turning timeline interpretation into an embodied spatial experience.  

We make three \textbf{contributions}: (\textbf{1}) Findings from a formative study (N=12) examining how people place event sequences in a physical environment when unconstrained by the convention of one-dimensional chronological ordering imposed by most visualization and data analysis tools; (\textbf{2}) A tablet-based augmented reality timeline visualization application that incorporates our study findings in its treatment of chronological representation, pacing, proxemic interaction, animation, and spatial context.
Finally, (\textbf{3}) we offer a reflection on how AR timeline visualization could be realized for personal storytelling and reflection as well as for biographical and historical storytelling in informal learning and GLAM environments (Galleries, Libraries, Archives, and Museums).



\section{Related Work}
\label{sec:rw}


Our work lies at the intersection of timeline visualization, personal / biographical visualization, and immersive analytics / storytelling. 

\subsection{Timeline Visualization}
\label{sec:rw:timeline}

For millennia, people have been using timelines to analyze and tell stories about patterns in event sequences~\cite{Rosenberg2010}.
Across the design space of timeline visualization~\cite{brehmer2016timelines}, myriad shapes, scales, and layouts have been put to use, as well as various techniques for interacting with and animating these representations. 
This design space illustrates trade-offs between readability and expressiveness. 
In practice, many timelines prioritize strict chronological representation, where spatial distance corresponds directly to elapsed time, and events are arranged along a single linear axis.



\bstart{Personal \& biographical (timeline) visualization}
Timelines often appear in the context of personal data.
Early work such as LifeLines~\cite{plaisant1996lifelines} demonstrated how timelines can summarize personal medical histories. 
While many instances of personal visualization are clinically useful, the study of personal visualization also examines how visual representations support reflection, memory, narrative construction, and meaning-making around lived experience~\cite{li2010stage}. 
To this end, prior work has examined how subjective time, emotional salience, and life events can be externalized into visual form, often privileging meaning over precision (\eg~\cite{perin2017symmetry}).
Personal visualization tools also tend to afford rich annotation, such as by enabling people to attach context and meaning to temporal data~\cite{walker2015timenotes}, whereupon time is treated as a structure for narrative interpretation rather than as a mere independent quantitative variable.

Prior work has shown that when unconstrained by software, personal mental models of time~\cite{hammond2012time} and the idiosyncrasies of personal timeline visualization become apparent, such as when sketching timelines on paper~\cite{torna2011}.
This in turn inspired TimeSplines~\cite{offenwanger2023timesplines}, in which people could bind event sequences to personally-meaningful 2D shapes sketched on a tablet. 
In some cases, the sketched shapes served as mnemonic~\cite{twain1914make} figurative frames~\cite{byrne2019figurative} for event sequences, whereas in others, the shapes reflected latent and subjective quantitative relationships and bore resemblance to time curves~\cite{bach2015time} and connected scatterplots~\cite{Haroz2015}. 
We extend this work by examining unconstrained timeline visualization in three dimensions and the role that (familiar) physical surroundings play in their creation.

\subsection{Immersive \& Embodied (Data) Storytelling} 
\label{sec:rw:immmersive}

Immersive storytelling explores how embodied interaction can support narrative understanding, often manifesting in augmented (AR) and virtual reality (VR). 
Prior work often distinguishes between egocentric experiences, which places the viewer at the center of a narrative, and exocentric perspectives, where people are detached from the spatial relationships being observed.
In our work, we primarily focus on egocentric immersive perspectives on time.
Two recent projects are exemplary of personal storytelling in VR environments: Tangible Moments~\cite{khan2025tangiblemoments} shows how embodied interaction can anchor autobiographical memories, highlighting the role of spatial context in recall, while MomentsVR~\cite{chowdhury2025momentsvr} focuses on experiential recall through immersive reconstruction of past events.

The design space of immersive \textit{data storytelling} experiences is growing. 
For example, the \textit{Wall Street Journal}’s ``Ride the NASDAQ'' interactive (initially developed for Google's Cardboard VR~\cite{cardboard}) presents financial data as a temporal roller-coaster experienced from a first-person perspective \cite{Pyle2014}. 
More recently, Jain~\etal's Strollytelling~\cite{jain2025strollytelling} demonstrates how walking and bodily movement can structure narrative progression in immersive environments. 
In our work, we focus on immersive storytelling in augmented reality (AR)~\cite{morgan2024storytelling}.
AR applications such as Flow Immersive~\cite{dibenigno2021flow} and research projects such as MARVisT~\cite{zhu2019marvist} and VisTellAR~\cite{tong2024vistellar} each demonstrate the creation and presentation of data stories in mixed reality, situated virtual representations of data into physical environments.
While these systems explore the narrative presentation of data in space, they prescribe narrative structures and perspectives. 
Less is known about how people would conceptualize narrative structures when creating stories about temporal data in the space around them.
Accordingly, we examine how participants assume egocentric and exocentric perspectives when constructing timelines in 3D space.

Beyond AR, immersive data storytelling also draws on embodied interaction with data in site-specific settings, where meaning is constructed through movement and physical engagement~\cite{cafaro2026data}. 
For example, Rodighiero~\etal~\cite{rodighiero2022surprise} demonstrated how full-body movement captured via webcam can drive the navigation of a visual browsing interface in a museum setting, while Perovich and Zizzi~\cite{perovich2024feeling} explored how the somatic practices of dancers can reflect a more affective and experiential understanding of data \cite{perovich2024feeling}. 
Together, these approaches emphasize that making sense of data is not purely visual, but can be deeply embodied, motivating our choreographic metaphor for describing how people construct temporal narratives while moving through space.







\bstart{Immersive timelines}
Three-dimensional representations of event sequences are uncommon in practice, with Tiki-Toki's~\cite{TikiToki} 3D timeline creator, the spiraling USGS timeline of geological epochs~\cite{Graham2008}, and the spiraling Ross school curriculum timeline~\cite{Ross2015} as rare exceptions. 
As for prior research, Fouch\'{e}~\etal's design space for immersive exploration of time-varying spatial 3D data~\cite{fouche2022timeline} identifies three important dimensions for visualizing time in AR/VR settings, namely spatial layout, scale, and navigation.
More recently, Vu~\etal~\cite{vu2025walking}~presented an application in which a person wears an AR-capable head-mounted display while interacting with a tablet to create and navigate circular and branching timelines. 
Although this research demonstrates the feasibility of immersive timelines, we note that they prescribe structures for time (\eg~linear, circular, branching), do not explicitly associate representations of events with the physical environment, and in the case of the latter project, require a specific combination of hardware devices. 
In contrast, our work offers insight into how people might invent or adapt timeline structures when given greater representational freedom.


\begin{figure*}[h!]
    \centering
    \includegraphics[width=\textwidth]{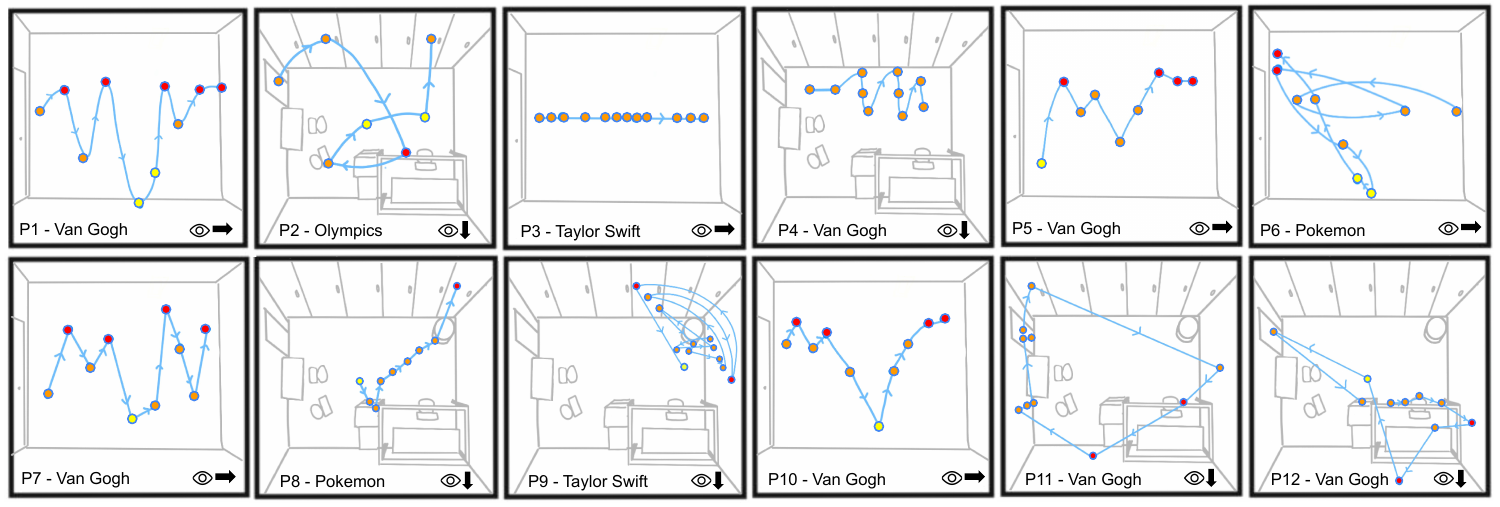}
    \caption{These timeline sketches show participants' placement of sticky notes in Part 1 of our formative study. Icons in the bottom right of each image indicate whether the diagram represents an eye-level perspective ($\rightarrow$) or an overhead perspective ($\downarrow$), while the \revision[luminance]{hue} of each point indicates its \revision[distance from the point of view of the sketch (darker is closer)]{height, with yellow indicating positions closer to the ground and red indicating positions closer to the ceiling}.}
    \label{fig:results-part1}
\end{figure*}
\section{Formative Study: Arranging Events in Space}

To better understand how people arrange sequences of events in their physical surroundings, we conducted a formative \revision[]{observational} study.
We had two constructs of interest: the effect of \textit{data familiarity}, or whether the data was autobiographical in nature; and the effect of \textit{place familiarity}, or whether the spatial arrangement of events is affected by meaningful surroundings. 
As \textbf{research questions}, we asked: (\textbf{RQ1}) whether and how idiosyncratic visual representations of time apparent in two-dimensional sketches~\cite{offenwanger2023timesplines} extend to three dimensions; as well as (\textbf{RQ2}) whether objects and elements in one's environment impact these representations, similar to how figurative illustrations in sketches can be integral to drawn timelines.  
The University of Waterloo research ethics board approved this study.

\bstart{Participants}
We recruited twelve participants from our campus and its surrounding community via multiple online channels and print \revision[poster advertising]{posters}. 
We had no exclusion criteria \revision[for participation]{} apart from being at least 18 years of age, and similar to the timeline sketching study of Torna~\etal~\cite{torna2011}, we assumed no technical expertise.

\revision[Overall, the group demographics reflected multiple gender identities as well as range of life experiences and fields of study.]{Seven participants used he/him pronouns, three used she/her pronouns, and two participants opted not to disclose their gender identity. Five were undergraduate students and seven were graduate students, with fields of study including computer science (5), mathematics (3), civil engineering, nursing, environment, and psychology. At the time of the study, two were between the ages of 18 and 20, seven were between 21 and 30, and three were between 30 and 40. Finally, participants reported a range of prior experience with extended reality technology, with three participants claiming no prior experience, two describing prior experience with mobile augmented reality games (specifically \textit{Pok\'emon Go}), and one claiming extensive virtual reality experience; the remaining reported minimal prior experience, such as participation in academic studies.}

We remunerated participants with a multi-retailer gift card valued between \$20 and \$30 CAD, depending on whether they completed both parts of the study.

\bstart{Study introduction} 
After acquiring informed consent, participants completed a brief questionnaire collecting demographic information including their age, gender, education level\revision[, and occupation]{field of study or occupation, and experience with extended reality technology}. 
The study administrator then introduced the study goals, stating that our findings could inform the development of augmented reality (AR) applications.    
They also introduced the study's focus on event sequence data and specifically historical and biographical events.
To support this briefing, they referred to a brief slide presentation containing examples of AR applications and alternative timeline representations of event sequences, as well as examples of augmented reality applications; we include these slides as supplemental material.
The intent of this introduction was to prime participants with respect to spatial thinking about time without prescribing specific metaphors or event placement strategies. 

\bstart{Materials}
Our study incorporated familiar low-fidelity materials (\ie~pen and sticky notes) to surface participants’ natural spatial reasoning without constraining interaction through existing visualization tools. We first arranged an example set of sticky notes around the study room. Each sticky note contained a date–event label pair, illustrating how temporal data could be externalized spatially without imposing a fixed coordinate system.

\bstart{Part 1: Arranging event sequences in an unfamiliar space} 
For the first task, participants were asked to select one of four provided datasets that we deemed to have historical or cultural significance, spanning eras and time spans: the biography of Vincent van Gogh, highlights from the 2010 Olympic games, the history of the Pok\'{e}mon franchise, and the release years of Taylor Swift's albums.
\revision[]{Each dataset consisted of roughly a dozen events,} recorded as a stack of sticky notes, participants freely placed these events around a controlled indoor environment to form a spatial timeline. They did so while filming their placement from a first-person perspective, holding a phone in their non-dominant hand.
This task functioned as an elicitation activity, allowing participants to reveal their underlying spatial metaphors for time.
During placement, we asked participants to think aloud, explaining their reasoning for each decision, including how they interpreted temporal ordering, distance, orientation, and alignment relative to the room and to previously placed events. The administrator observed and documented placement decisions through sketches and field notes, while participants’ movements and utterances were recorded. This procedure enabled analysis from both egocentric and exocentric frames of reference.

Following the placement of culturally significant data, we gave participants a blank set of sticky notes and asked them to create a personally meaningful timeline by writing down approximate dates and short labels for about a dozen events of their choosing (\eg~ vacations, achievements, routines, or relocations). Participants then repeated the placement task in the same (unfamiliar) study room environment, allowing for a direct comparison between personal and impersonal data within a consistent spatial context.

\bstart{Part 2: Personal events in a familiar space}
We invited participants to repeat the personal timeline placement task within a familiar environment of their own choosing, such as their dorm room or their kitchen. 
This phase was designed as a contextual inquiry, situating the elicitation task within a space rich with personal landmarks and semantic associations. 
\revision[]{Because we opted to conduct a qualitative observation-based formative study rather than an experiment with counter-balanced conditions, we did not ask participants to repeat the placement of historically or culturally significant events in familiar spaces. In addition to imposing an additional time burden on participants, we doubted the external validity of annotating a personal space with impersonal data.}

Participants joined a remote video call with the study administrator using their mobile device. They toggled their phone's primary camera and provided a brief visual walkthrough of their space, allowing the administrator to make note of salient landmarks (\eg~furniture, windows, artwork) and spatial constraints. Participants then placed their set of personal event sticky notes around the space, again filming their placement with their mobile device.
As in Part 1, we asked participants to think aloud and describe their rationale for placing events near objects and features in their surroundings and the personal memories associated with these objects. 

\bstart{Retrospective interview}
After completing the placement tasks, participants engaged in a semi-structured retrospective interview. We asked participants to contrast their experiences across conditions (personal vs. impersonal data; familiar vs. unfamiliar space), the perceived affordances and limitations of the materials used, and participants’ expectations for how AR applications might support or augment similar storytelling and reflection activities.

Overall, completing all parts of the study required roughly 90 minutes of participants' time.

\bstart{Analysis}
\revision[Our]{The goal of our} analysis \revision[of the sketches and video recordings produced during the study focused on identifying]{was to identify} spatial metaphors and placement strategies across conditions (\textbf{RQ1}), with a focus on the differences between general and personally meaningful data, as well as between familiar and unfamiliar spaces (\textbf{RQ2}). 
Throughout this analysis, we referred to participants' think-aloud utterances and retrospective interview responses to better understand their rationale for placing the sticky notes where they did.
\revision[]{The first two authors conducted this analysis in two phases, with the second author's annotated screen capture images, notes, and sketches serving as an input to the first author's systematic shape and chronology analysis, along with the video recordings. All authors reviewed the output of these phases, reaching the consensus presented below.}

\section{Results: Shape, Chronology \& Situatedness}

\revision[]{While all four sample datasets appear in our observations of participants arranging culturally significant datasets, they are not equally represented, suggesting a range of familiarity and interest. This is reflected in how seven of our twelve participants selected the van Gogh biography dataset in the first part of their study session.}

\begin{figure}[!h]
    \centering
    \includegraphics[width=0.5\textwidth]{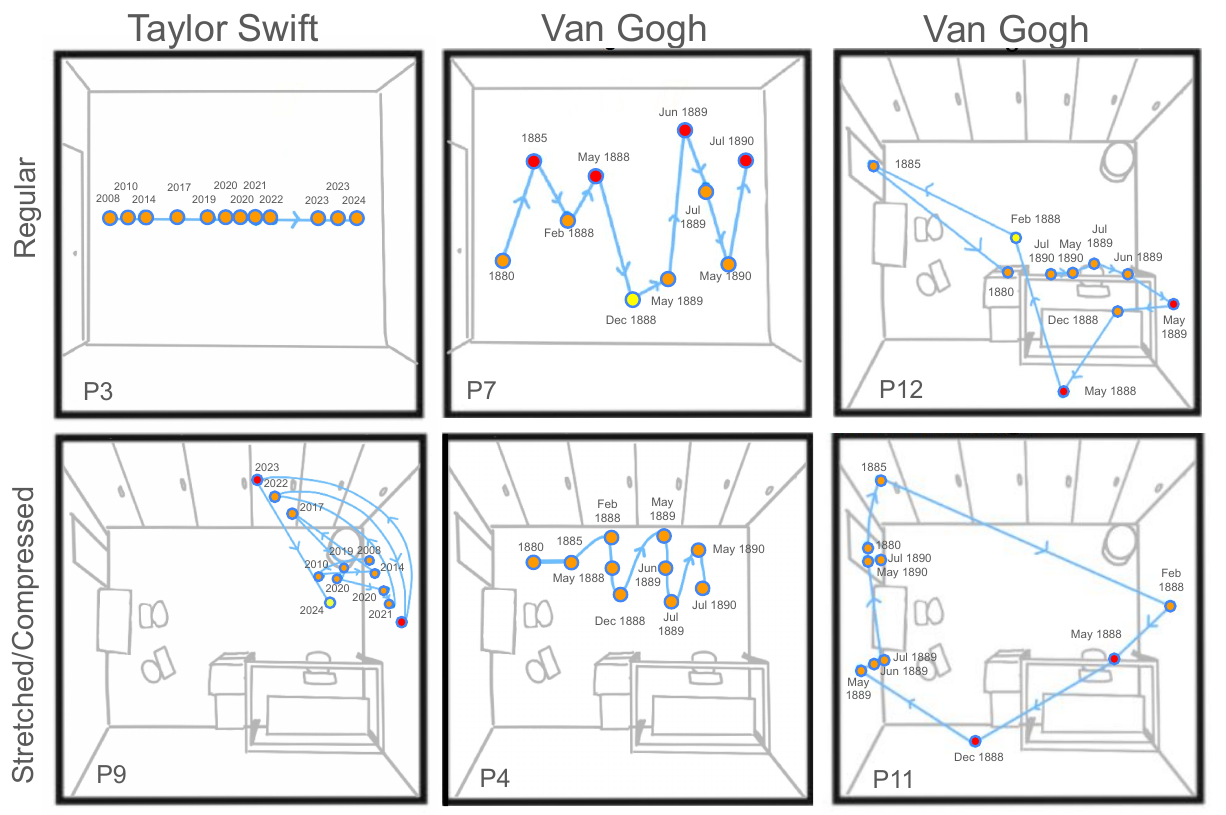}
    \caption{We annotate a subset of sketches from \autoref{fig:results-part1} to contrast sequential and chronological regularity with chronological compression and expansion between events.}
    \label{fig:results-chronology}
\end{figure}

\begin{figure}[!h]
    \centering
    \includegraphics[width=0.5\textwidth]{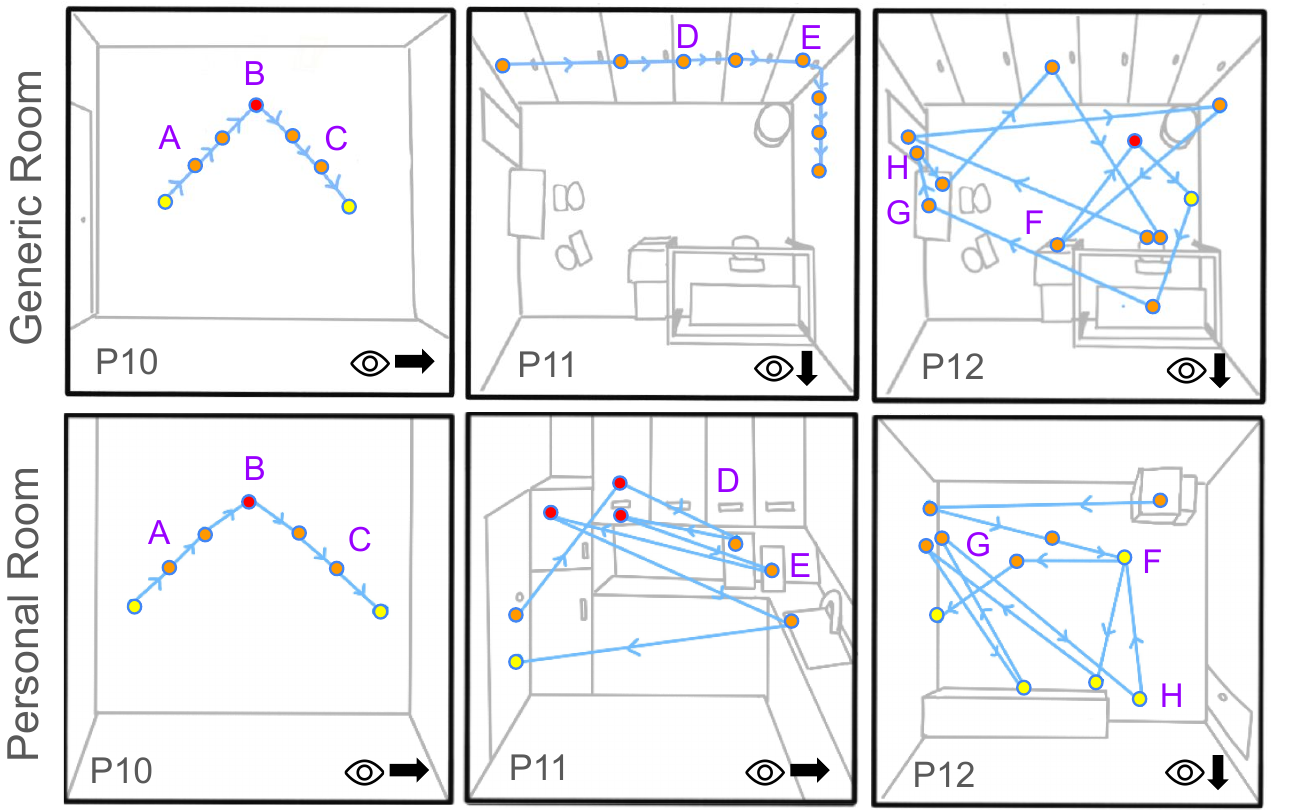}
    \caption{Sketches contrasting the placement of personal events by three participants (P10,P11,P12) between an unfamiliar room and one of their own choosing, with matching annotations (A-H) corresponding to the same events.}
    \label{fig:results-space}
\end{figure}

\bstart{Shape analysis}
In examining all of our sketched timelines (Figures~\ref{fig:results-part1}---\ref{fig:results-space}), shape emerged as a primary expressive channel for encoding meaning beyond chronology.
The variety of timeline geometries is notable and includes linear, looping, and stacked shapes. 
\revision[]{With respect to linear shapes,} about half of the participants (P1, P3, P5, P6, P7, P10) used a two-dimensional plane within the room as a pair of axes, encoding emotional valence or subjective intensity along one dimension and time along the other, indicating that planar representations of time remain prevalent even when given the freedom to place events anywhere within a room. 
Three-dimensional \revision[shapes]{loops and stacks} appeared when participants sought to express simultaneity, complexity, or layered meaning. 
Examples of this strategy include stacking events that happened in the same time period (P3) \revision[]{(while maintaining a macro-level linear progression of time)}, using depth for years and lateral movement for months (P4) and attaching events to furniture or architectural features (P4, P5, P11, P12). 
To summarize, our shape analysis reveals that \textit{the relative positioning of events in space reflects not only chronology but also event distribution, the affective attributes of events, and the physical affordances of the environment} (\textbf{RQ1}).

\bstart{Chronology analysis}
Although our participants acknowledged the chronology of events, the distances between participants' placed events seldom approximated accurate chronological distances. 
Time was therefore stretched or compressed; we contrast some notable cases in ~\autoref{fig:results-chronology}.
Participants commonly decoupled spatial distance from absolute chronological time and instead 
P7, for example, employed equidistant \revision[sequential]{horizontal} spacing in their placement reflecting van Gogh's biography, while P3 and P9 grouped events ordinally by year or life phase, rather than respecting precise chronological distances. 
P2 and P12 both made remarks to the effect of events mattering more than time itself, reflecting their treatment of chronology as a secondary ordering mechanism.
Periods with minimal emotional or narrative significance were frequently compressed, producing large chronological gaps occupying small spatial regions (\eg~P4, P7), whereas dense or meaningful life phases expanded disproportionately in space (\eg~P11). 
Altogether, these observations suggest that \textit{the combination of sequential / ordinal placement and the perceived importance of events can result in notable chronological compression and expansion} (\textbf{RQ1}). 

\bstart{Analyzing the role of room features and objects} 
Across both parts of the study, participants leveraged semantic associations with the physical environment to anchor data, using architectural features and everyday objects as symbolic reference points rather than neutral containers. 
Doors marked beginnings, transitions, or goals (P2, P7, P9, P11, P12), while walls served as narrative backbones, particularly in unfamiliar spaces where flat and unadorned surfaces supported linear organization (P3, P7, P10). 
Windows and thresholds implicitly represented outward movement, change, or external influence (P5, P9). 
Furniture and smaller objects functioned as mnemonic anchors; for example, P8 and P9 placed events with a negative emotional valence near a trash can (P8, P9), while P2, P5, and P11 placed events associated with routines or habits next to a laptop, a musical instrument, and a coffee machine, respectively.
Participants also placed events next to objects with semantically-related material qualities like colour and texture, such as in the use of a yellow surface for a yellow house (P11). 
Overall, \textit{semantic event anchoring was more apparent with personally meaningful datasets, where room features and objects were more than mere visual landmarks; they also carried symbolic significance} (\textbf{RQ2}).

\bstart{Analyzing the role of familiar space}
All but three of the original twelve participants (P1, P6, P8) opted to participate in Part 2 of the study.
A clear distinction in placement strategy emerged in how participants represented their own personal events versus the event datasets we provided, and this was also mediated by the familiarity of the surrounding space \revision[]{(\textbf{RQ2})}. 
In the unfamiliar space of the study room (\autoref{fig:teaser}), the resulting timeline shapes were more likely be linear in shape and aligned with walls, respecting the conventions of left-to-right temporal progression and vertical mappings of emotional valence (\eg~P3, P7, P10 in \autoref{fig:results-part1}). 
In contrast, personal datasets elicited more creativity, reflected in participants’ willingness to distort time, incorporate multisensory cues such as music or lighting, and adopt a more overt egocentric perspective rather than an exocentric perspective (P12). 
Additionally, familiar personal spaces elicited object-anchored, room-specific, and more embodied placements that drew on memory-rich landmarks, such as specific books and accessories (P5).
Two notable examples are the cases of P11 and P12 (\autoref{fig:results-space}).
P11 placed events in their kitchen reflecting their growing appreciation of coffee over the span of nine years centering around their espresso machine, while P12 arranged events in their bedroom relating to their theological journey over the course of ten years, placing events near personally significant objects taken home from religious events.
However, we still encountered planar event placements in familiar environments, such as how P10 arranged major travel events spanning six years on their bedroom wall. 
Despite exceptions like this, multiple participants made comments indicating that \textit{the same set of events would be arranged quite differently depending on spatial context} (\textbf{RQ2}).

\section{AR Timeline Prototype}
\label{sec:system}

Informed by our analysis in service of \textbf{RQ1} and \textbf{RQ2}, we designed and implemented a tablet-based augmented reality (AR) application that supports the drawing and viewing of timelines that reflect how people place events in physical space.
It also reflects participants' statements reflecting how they might share their timeline stories with other people, including what guidance or narrative framing they might provide to their viewers. 
Unexpectedly, the act of sketching the event sequences (leading to Figures~\ref{fig:results-part1}---\ref{fig:results-space}) brought us to a choreographic metaphor for drawing timelines in augmented reality. 
These sketches reminded us of the visual language of dance choreography, and given how affect-based event placement and shape variation are exhibited across these sketches, we deemed the metaphor to be an appropriate way to frame the application and the discussion that follows in \cref{sec:discussion}.

\begin{figure*}[!h]
    \centering
    \includegraphics[width=\textwidth]{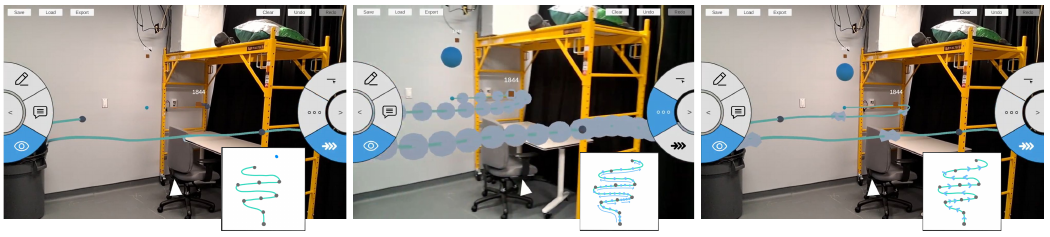}
    \caption{Our prototype has three viewing modes corresponding to different treatments of chronology, where visual connections between events appear progressively \revision[]{(represented abstractly in the inset diagrams)}: no chronological encoding (left), chronological distance encoded as the speed of animated particles between events (middle), and chronological distance encoded via textured glyphs along line segments (right). In each mode, viewers are guided to the next event (a spherical blue waypoint) with a compass (a white chevron).}
    \label{fig:system-overview}
\end{figure*}

\bstart{Scope \& design choices} 
A key difference between our study participants' experiences and our sketches is that the former assumed an egocentric perspective, whereas our ``choreographic'' sketches are exocentric. 
We ultimately decided that, like in our formative study protocol, our application would assume an egocentric perspective for both timeline drawing and viewing. 
Viewers therefore retrace the motion of timeline creators to view an entire event sequence, with a pacing determined by the creators.

Another observation from the sketches is that when events are strung together with line segments, the result can be visually cluttered from an exocentric perspective. 
We discovered during initial prototyping that this clutter remains apparent when we assume an egocentric perspective.
To remedy this, our application incorporates proxemic interaction to progressively reveal the next segment of the timeline after a viewer visits an event location.  

Finally, while we acknowledge the importance of juxtaposing events with objects and room features in the formative study results, our prototype application does not employ any object or feature detection in the scene.
As a result, events can be juxtaposed with objects or features but are not bound to them; if \revision[the]{}these objects or features move or change shape, the events and the timeline by extension are unaffected. 


\subsection{Drawing \& Viewing Interface}
We selected a tablet form factor for both timeline creation and viewing for several reasons. 
First, tablets with support for AR functionality are  broadly available, more affordable than head-mounted displays, and have a larger screen than a phone.
Second, a person can walk around relatively easily while holding a tablet with one or both hands.
Finally, a tablet affords both touch and stylus input.

The bimanual AR tablet interface supports dual thumb or optional thumb + pen interaction~\cite{pfeuffer2017thumb+}. 
One can control the interface with thumbs only via the lateral radial buttons on either side of the screen (\autoref{fig:system-overview}). 
This design follows prior research showing that adaptable, thumb-reachable radial widget layouts improve comfort, efficiency, and accuracy on tablets \cite{perelman2024dual}. 
For higher precision when drawing and manipulating the shape of a timeline, one can use a stylus, following the Thumb + Pen interaction paradigm~\cite{pfeuffer2017thumb+} where the stylus handles fine input while the thumb selects controls.
The left radial menu allows for mode switching between timeline drawing, annotating (\autoref{fig:teaser}F), and viewing (\autoref{fig:system-overview}), while the right radial menu's widgets are dependent on the mode. 

\subsection{Timeline Creation Scenario}

To illustrate the authoring workflow, we describe a scenario in which Alex, a student studying the history of computing, draws a biographical timeline of Ada Lovelace (also demonstrated in the supplemental video).
Alex has a Google spreadsheet containing dates and associated event descriptions, which the application imports. 

Holding their tablet, Alex walks to sketch the overall shape of the timeline. 
As they move across the room, their motion path is translated to a path element representing the progression of time. 
Once drawn, Alex toggles a path editing widget and uses a stylus to precisely refine the path shape.
Next, Alex places events along the path, representing major moments in Lovelace’s life, such as her early mathematical education and her collaboration with Charles Babbage. 
Each event snaps to a selected location on the path and is repositioned via dragging. 
To denote chronology, Alex also adds visual waypoints between events that serve as visual indicators of time, akin to year markers along the horizontal span of a line chart.  
Alex alternates between editing the path shape and iteratively adjusting the spacing between events and waypoints, clustering important subsequences and expanding the distance between phases of Lovelace’s life.
Periodically, Alex toggles transformation widgets to translate, scale, and rotate the entire timeline in the room.
Like any drawing tool, Alex occasionally needs to undo or redo their edits.
Once satisfied, Alex saves the timeline so it can later be revisited, edited, or viewed by someone else.





\subsection{Timeline Viewing Experience(s)}

\bstart{Orientation and proxemic interaction}
The application reveals a timeline one event at a time, with a compass guidance system directing viewers to the next event. 
Each event appears based on viewer proximity, manifesting as a floating event description above the event marker. 
The application then reveals the next path segment after a dwell time set by timeline creator, with a visual progress indicator shown under the event description. 
This allows creators to control the pacing of the timeline viewing experience, such as by asking viewers to linger and reflect at important events. 

\bstart{Three treatments of chronology}
We developed three viewing modes to reflect the \revision[various ways by which chronology manifested]{compression and expansion of chronology manifesting} in our formative study results (\autoref{fig:system-overview}).
\revision[]{These treatments represent points in a larger design space for visual encodings of chronology in three dimensions, a design space that invites further exploration.}

\istart{No encoding}
If chronology is of minor importance to the intended narrative, this mode (\autoref{fig:system-overview}-left) merely connects sequential events with a path.
While chronology is not explicitly encoded in this mode, chronological waypoint markers (described above) can still provide viewers with some sense of chronology.

\istart{Animated encoding}
Motivated by the ways in which people compressed or expanded chronology, this mode (\autoref{fig:system-overview}-middle) encodes the chronological distance between events via animate particles moving along the forward direction of time.
Faster particles 
indicate compressed or less significant intervals, while slower particles indicate expanded and meaningful passages of time. 

\istart{Textured encoding} 
Realizing that animated particles could be visually distracting, the third mode (\autoref{fig:system-overview}-right) employs a texture of chevron shapes along the path pointing in the direction of time.
As with animation, denser textures denote compressed intervals while sparser segments indicate expanded time. 



\subsection{Implementation}
We implemented the prototype in Unity 6~\cite{unity} with Android ARCore~\cite{arcore} for device tracking and spatial anchoring.
We deployed it on a Samsung Galaxy Tab FE10 running Android OS.
Its source code is available at \href{https://github.coim/ubixgroup/ar-timeline-authoring-tool}{github.com/ubixgroup/ar-timeline-authoring-tool}.


\section{Discussion}
\label{sec:discussion}

The formative study findings highlight the diverse ways by which people conceptualize and represent time, aligning with prior work in information design~\cite{torna2011} and time perception~\cite{hammond2012time}.
We add to this knowledge by explicitly incorporating the dimension of place familiarity. 
We observed that participants exploited the affordances of three-dimensional space to represent time when given the opportunity, particularly in familiar environments. 

Our research questions were also motivated by the growing potential of spatial computing technology, and our analysis is intended to inform the design of future augmented reality (AR) experiences.
We believe our choreographic metaphor framing this analysis to be actionable in this regard, whereupon content creators can direct movement through space to understand time by varying direction and pacing. 
However, other spatial metaphors are also worth exploring in this context, including those from cinematography~\cite{amini2015understanding}, interior design~\cite{brehmerdata}, and the practice of Feng Shui~\cite{han2025feng}.

Our participants described whether and how they might share their timeline stories with others, which motivated the distinct creation and viewing modes of our prototype application.
We see creators taking on the role of choreographers, considering how viewers would move along their timelines, where attention should be directed, and how long a viewer might linger at particular moments. 
Applications descending from our prototype could integrate prior techniques in site-specific AR visualization (\eg~\cite{tong2024vistellar,zhu2019marvist}), offering a rich medium for asynchronous interpersonal storytelling.
It is within informal educational environments such as GLAM spaces (Galleries, Libraries, Archives, and Museums) where this medium could be particularly valuable, with artifacts augmented with additional visual representations and personalized navigational assistance guiding viewers to the next artifact on the timeline. 

Our work also elicits new questions with respect to the utility of AR timeline visualization beyond interpersonal storytelling. 
One such question asks about the memorability of situated AR timelines in immersive analytics scenarios~\cite{liu2024investigating, hurter2024memory}.
Another question is how choreographic timeline visualization could be applied to instructional guidance in AR~\cite{sadprasid2024improving}.
Finally, we ask whether our approach could be integrated into one's own personal data tracking and reflection practice.
AR timelines depicting how one spends their time within a home or workplace could assist in personal time management and memory care. 
These spatial timelines would likely manifest as small-scale analogues to Otten~\etal's~\textit{shifted maps}~\cite{otten2018shifted}, which illustrate skewed personal geographies.
Moreover, these timelines would likely reinforce the notion that time is not fungible~\cite{odell2024saving}, and that regular chronological depictions of time divorced from where that time is spent are poor tools for reflection.

\bstart{Study limitations}
While our study provided valuable insights into how participants arrange events in physical space, several limitations should be acknowledged. 
First, connections between events were not explicitly visualized, so the relationships between them remained implicit from participants' perspectives; we entertained the idea of asking participants to join their events with segments string or yarn with varying levels of slackness to encode chronology, but ultimately we decided against this due to increased creation time and obvious safety risks.
Second, due to the use of sticky notes, participants could not place events in mid-air, so we instead instructed them to place them on the ground and express verbally how high off of the ground they intended the event to appear. 
Third, in the remote second part of the study, there was unavoidable ambiguity in the analysis of event placements due to the research team's unfamiliarity with participants' personal locations, seen solely through participants' first-person perspective video. 
\revision[]{Finally, three of our twelve participants opted not to participate in the second part of the study, in which we asked them to place personal events in a familiar space. 
Attrition is expected and difficult to avoid in studies with multiple sessions and locations, and we attribute some of this attrition to the logistical complexity of study conducted across two sessions and in two locations. Another explanation for this attrition may be privacy, as we asked participants to share video of a familiar space, which we anticipated being a room in their home.
To address this in future studies, we could opt for asynchronous video diary recording rather than synchronous video observation, or we could consider other sources of incentivization relevant to documenting and sharing personal timelines.} 

\bstart{Future application development}
We see several ways to modify or extend the prototype application so that it might reflect more of the study findings.
First, integrating scene understanding, including the identification and tracking of objects and features in the environment, could allow for dynamic timeline shape manipulation via the movement of objects bound to events. 
Scene understanding could also be useful for the responsive retargeting of timelines between rooms that vary with respect to their dimensions but contain similar objects (\eg kitchens containing common utensils and appliances) or share similar features (\eg doors, windows, chairs, tables).
Second, additional representations of chronology beyond those described in \cref{sec:system} and shown in \autoref{fig:system-overview} are warranted. 
Whether unadorned, animated, or textured, events were connected with linear path segments. 
One promising alternative is a physical analogy of string or yarn that we briefly entertained for the formative study, in which long passages of time between events could hang slackly, perhaps even coiling on the ground, while short passages of time could be appear tight. 
Third, we might revisit the creation workflow, which is currently similar to TimeSplines~\cite{offenwanger2023timesplines} in that the path of time is drawn and manipulated before events are bound to it.
Alternatively, events could be placed before connecting them visually.
\revision[]{Placing events could also be done more flexibly that at current, such as by projecting events at different depths in the room, rather than walking and moving the tablet to a desired position during timeline creation.}
Finally, reflecting the exocentric perspective of the sketches in Figures~\ref{fig:results-part1}---\ref{fig:results-space} or the inset diagrams of \cref{fig:system-overview}, we ask whether and how the application should expose this perspective when creating or viewing timelines. 

\bstart{Future research}
An immediate next step is to conduct a replication of the procedure from our formative study with the prototype application.
Such a study would allow us to contrast how the shape, chronology, and situatedness of events vary when they could be placed anywhere within reach, and how visually connecting these events with our different treatments of chronology inform additional event manipulation. 
Re-engaging our formative study participants and recruiting new ones while explicitly targeting different demographics  could produce an even richer set of contrasts. 
\revision[]{Finally, the realization of augmented reality timelines revealed via proxemic interaction invites new research focusing on accessibility, particularly if they are to be situated in GLAM spaces attended by those with a different abilities with respect to vision and mobility.}  


\section{Conclusion}
This work investigated how event sequences can be arranged in space and visualized as timelines in augmented reality (AR).
Through a formative study, we observed the idiosyncrasies of event placement, in which participants navigated both unfamiliar and familiar spaces to place sticky notes describing historical and biographical events, including those drawn from their own personal stories.
The rationale for their placement and the patterns that emerged from our analysis led us to a choreographic metaphor for AR timeline creation and viewing.
This metaphor incorporates the shape and pacing of movement and manifests in our prototype application via proxemic progressive reveal and suggestive visual encodings reflecting the passage of time. 
Overall, our research suggests the potential of spatial timelines in AR as a powerful medium for storytelling and reflection that emphasizes movement, rhythm, and embodied interaction as core elements of narrative experience.


\bibliographystyle{ACM-Reference-Format}
\bibliography{references}

@string{TVCG = "IEEE Trans. Visualization \& Computer Graphics (TVCG)"}

@STRING{CHI = {ACM Conf. Human Factors in Computing Systems (CHI)}}

@string{VRST = {ACM Symp. Virtual Reality Software \& Technology (VRST)}}

@string{VIS = {IEEE Conf. Visualization \& Visual Analytics (VIS)}}

@string{TEI = {ACM Intl. Conf. Tangible, Embedded, and Embodied Interaction (TEI)}}

@string{IDJ = {Information Design Journal}}

@article{brehmer2016timelines,
title={Timelines revisited: A design space and considerations for expressive storytelling},
author={Brehmer, Matthew and Lee, Bongshin and Bach, Benjamin and Riche, Nathalie Henry and Munzner, Tamara},
journal=TVCG,
volume={23},
number={9},
year={2016},
doi={10.1109/TVCG.2016.2614803}
}

@inproceedings{fouche2022timeline,
title={Timeline design space for immersive exploration of time-varying spatial 3d data},
author={Fouch{\'e}, Gwendal and Argelaguet Sanz, Ferran and Faure, Emmanuel and Kervrann, Charles},
booktitle=VRST,
year={2022},
doi={10.1145/3562939.3565612}
}

@article{bach2015time,
title={Time curves: Folding time to visualize patterns of temporal evolution in data},
author={Bach, Benjamin and Shi, Conglei and Heulot, Nicolas and Madhyastha, Tara and Grabowski, Tom and Dragicevic, Pierre},
journal=TVCG,
volume={22},
number={1},
year={2015},
doi={10.1109/TVCG.2015.2467851}
}

@inproceedings{jain2025strollytelling,
title={Strollytelling: Coupling animation with physical locomotion to explore immersive data stories},
author={Jain, Radhika Pankaj and Drogemuller, Adam and Satriadi, Kadek Ananta and Smith, Ross and Cunningham, Andrew},
booktitle=CHI,
year={2025},
doi={10.1145/3706598.3713132}
}

@article{zhu2019marvist,
title={{MARVisT}: Authoring glyph-based visualization in mobile augmented reality},
author={Zhu-Tian, Chen and Su, Yijia and Wang, Yifang and Wang, Qianwen and Qu, Huamin and Wu, Yingcai},
journal=TVCG,
volume={26},
number={8},
year={2019},
doi={10.1109/TVCG.2019.2892415}
}

@article{tong2024vistellar,
title={{VisTellAR}: Embedding data visualization to short-form videos using mobile augmented reality},
author={Tong, Wai and Shigyo, Kento and Yuan, Lin-Ping and Fan, Mingming and Pong, Ting-Chuen and Qu, Huamin and Xia, Meng},
journal=TVCG,
volume={31},
number={3},
year={2024},
doi={10.1109/TVCG.2024.3372104}
}

@inproceedings{vu2025walking,
title={Walking through time: A hybrid immersive system for spatial and expressive timeline authoring},
author={Vu, Veronica and Rai, Yogesh and Chung, Haeyong},
booktitle=VIS,
year={2025},
doi={10.1109/VIS60296.2025.00076}
}

@inproceedings{plaisant1996lifelines,
title={{LifeLines}: visualizing personal histories},
author={Plaisant, Catherine and Milash, Brett and Rose, Anne and Widoff, Seth and Shneiderman, Ben},
booktitle=CHI,
year={1996},
doi={10.1145/238386.238493}
}

@inproceedings{li2010stage,
title={A stage-based model of personal informatics systems},
author={Li, Ian and Dey, Anind and Forlizzi, Jodi},
booktitle=CHI,
year={2010},
doi={10.1145/1753326.1753409}
}

@article{walker2015timenotes,
title={Timenotes: A study on effective chart visualization and interaction techniques for time-series data},
author={Walker, James and Borgo, Rita and Jones, Mark W},
journal=TVCG,
volume={22},
number={1},
year={2015},
doi={10.1109/TVCG.2015.2467751}
}

@inproceedings{khan2025tangiblemoments,
title={{TangibleMoments}: Embedding XR memories onto physical objects},
author={Khan, Omar and Ahmed, Zaid and Nam, Hyeongil and Kim, Kangsoo},
booktitle={IEEE Conf. Virtual Reality and 3D User Interfaces (VRW)},
year={2025},
doi={10.1109/VRW66409.2025.00227}
}

@inproceedings{chowdhury2025momentsvr,
title={{MomentsVR}: Accessible immersive spaces for authoring personal narration},
author={Chowdhury, Nabila and Offenwanger, Anna and Tsandilas, Theophanis and Chevalier, Fanny and Ahmed, Syed Ishtiaque},
booktitle={Extended Abstracts ACM Conf. Computer-Supported Cooperative Work and Social Computing (CSCW): Companion Publication},
year={2025},
doi={10.1145/3715070.3749270}
}

@article{Pyle2014,
author = {Kenny, Roger and Becker, Ana Asnes},
title = {Is the Nasdaq in Another Bubble?},
journal = {The Wall Street Journal},
year = {2015},
note = {\href{https://web.archive.org/web/20151106015312/https://graphics.wsj.com/3d-nasdaq/}{graphics.wsj.com/3d-nasdaq/}}
}

@book{hammond2012time,
title={Time Warped: Unlocking the Mysteries of Time Perception},
author={Hammond, Claudia},
year={2012},
publisher={House of Anansi}
}

@misc{arcore,
author={Google},
title={ARCore},
year={2018},
note={\href{https://developers.google.com/ar}{developers.google.com/ar}}
}

@misc{cardboard,
author={Google},
title={Cardboard},
year={2014},
note={\href{https://web.archive.org/web/20230506183532/https://arvr.google.com/cardboard/}{arvr.google.com/cardboard/}}
}

@misc{unity,
key={Unity Technologies},
title={Unity 6},
year={2025},
note={\href{https://unity.com/releases/unity-6}{unity.com/releases/unity-6}}
}

@article{offenwanger2023timesplines,
title={Timesplines: Sketch-based authoring of flexible and idiosyncratic timelines},
author={Offenwanger, Anna and Brehmer, Matthew and Chevalier, Fanny and Tsandilas, Theophanis},
journal=TVCG,
volume={30},
number={1},
year={2023},
doi={10.1109/TVCG.2023.3326520}
}

@incollection{torna2011,
title={Visualizing time},
author={Torna, Camilla},
editor = {Ast, Olga},
booktitle={Infinite Instances: Studies and Images of Time},
pages={42--51},
year={2011},
publisher={Mark Batty Publisher}
}

@article{rodighiero2022surprise,
title={Surprise machines: revealing Harvard Art museums’ image collection},
author={Rodighiero, Dario and Derry, Lins and Duhaime, Douglas and Kruguer, Jordan and Mueller, Maximilian C and Pietsch, Christopher and Schnapp, Jeffrey T and Steward, Jeff},
journal=IDJ,
volume={27},
number={1},
year={2022},
doi={10.1075/idj.22013.rod}
}

@inproceedings{perovich2024feeling,
title={Feeling data through movement: Designing somatic data experiences with dancers},
author={Perovich, Laura J and Zizzi, Nicole},
booktitle=TEI,
year={2024},
doi={10.1145/3623509.3633371}
}

@book{cafaro2026data,
title={Data through movement: Designing embodied human-data interaction for informal learning},
author={Cafaro, Francesco and Roberts, Jessica},
year={2026},
publisher={Springer Nature}
}

@inproceedings{pfeuffer2017thumb+,
title={Thumb+ pen interaction on tablets},
author={Pfeuffer, Ken and Hinckley, Ken and Pahud, Michel and Buxton, Bill},
booktitle=CHI,
doi={10.1145/3025453.3025567},
year={2017}
}

@article{perelman2024dual,
title={Dual-thumb ointing and command selection techniques for tablets: Enhancing distant interaction with large displays using a two-handed tablet},
author={Perelman, Gary and Dubois, Emmanuel and Serrano, Marcos},
journal={International Journal of Human-Computer Studies},
volume={184},
year={2024},
doi={10.1016/j.ijhcs.2023.103203}
}

@inproceedings{liu2024investigating,
title={Investigating the effects of physical landmarks on spatial memory for information visualisation in augmented reality},
author={Liu, Jiazhou and Satriadi, Kadek Ananta and Ens, Barrett and Dwyer, Tim},
booktitle={Intl. Symp. Mixed and Augmented Reality (ISMAR)},
year={2024},
doi={10.1109/ISMAR62088.2024.00043}
}

@book{morgan2024storytelling,
title={Storytelling for Spatial Computing and Mixed Reality: The Art of Augmenting Imagination},
author={Morgan, Rob},
year={2024},
publisher={CRC Press}
}

@article{hurter2024memory,
title={Memory recall for data visualizations in mixed reality, virtual reality, {3D} and {2D}},
author={Hurter, Christophe and Rogowitz, Bernice and Truong, Guillaume and Andry, Tiffany and Romat, Hugo and Gardy, Ludovic and Amini, Fereshteh and Riche, Nathalie Henry},
journal=TVCG,
volume={30},
number={10},
year={2024},
doi={10.1109/TVCG.2023.3336588}
}

@misc{TikiToki,
author = {{Webalon}},
key = {{Webalon}},
title = {{Tiki-Toki}},
year = {2010},
note = {\href{http://tiki-toki.com/}{tiki-toki.com}}
}

@misc{Ross2015,
author = {{Ross Institute} and {Moebio Labs}},
title = {Ross spiral curriculum},
year = {2015},
note = {\href{http://spiral.rosslearningsystem.org/spiral/}{spiral.rosslearningsystem.org/spiral}}
}

@book{Rosenberg2010,
author = {Rosenberg, Daniel and Grafton, Anthony},
title = {{Cartographies of Time: A History of the Timeline}},
publisher = {Princeton Architectural Press},
year = {2010}
}

@article{Haroz2015,
author = {Haroz, Steve and Kosara, Robert and Franconeri, Steven L.},
title = {The connected scatterplot for presenting paired time series},
journal = TVCG,     
doi = {10.1109/TVCG.2015.2502587},
year = {2016}, 
volume = {22}, 
number = {9}, 
}

@article{byrne2019figurative,
title={Figurative frames: A critical vocabulary for images in information visualization},
author={Byrne, Lydia and Angus, Daniel and Wiles, Janet},
journal={Information Visualization},
volume={18},
number={1},
year={2019},
doi={10.1177/1473871617724212}
}

@article{twain1914make,
author = {Twain, Mark},
title = {How to make history dates stick},
journal = {Harper's Monthly Magazine},
number = {775},
volume = {130},
year = {1914},
note = {\url{http://www.twainquotes.com/HistoryDates/HistoryDates.html}}
}

@inproceedings{perin2017symmetry,
title={The symmetry of my life: An autobiographical visualization},
author={Perin, Charles},
booktitle={IEEE Conf. Visualization \& Visual Analytics (VIS) Poster Proceedings},
year={2017},
note={\href{https://inria.hal.science/hal-01587944}{inria.hal.science/hal-01587944}}
}

@article{dibenigno2021flow,
title={{Flow Immersive}: A multiuser, multidimensional, multiplatform interactive {COVID-19} data visualization tool},
author={DiBenigno, Michael and Kosa, Mehmet and Johnson-Glenberg, Mina C},
journal={Frontiers in Psychology},
volume={12},
year={2021},
doi={10.3389/fpsyg.2021.661613}
}

@inproceedings{otten2018shifted,
title={{Shifted Maps}: Revealing spatio-temporal topologies in movement data},
author={Otten, Heike and Hildebrand, Lennart and Nagel, Till and D{\"o}rk, Marian and M{\"u}ller, Boris},
booktitle={IEEE VIS Arts Program (VISAP)},
year={2018},
doi={10.1109/VISAP45312.2018.9046054}
}

@inproceedings{sadprasid2024improving,
title={Improving video navigation for spatial task tutorials by spatially segmenting and situating how-to videos},
author={Sadprasid, Book and Gutwin, Carl and Bateman, Scott},
booktitle={ACM Symp. Spatial User Interaction (SUI)},
doi={10.1145/3677386.3682103},
year={2024}
}

@inproceedings{amini2015understanding,
title={Understanding data videos: Looking at narrative visualization through the cinematography lens},
author={Amini, Fereshteh and Henry Riche, Nathalie and Lee, Bongshin and Hurter, Christophe and Irani, Pourang},
booktitle=CHI,
doi={10.1145/2702123.2702431},
year={2015}
}

@inproceedings{han2025feng,
title={The {Feng Shui} of visualization: Design the path to success and good fortune},
author={Han, Chang and McNutt, Andrew},
note={Proc. IEEE VIS alt.VIS Workshop},
doi={10.48550/arXiv.2510.00344},
year={2025}
}

@inproceedings{brehmerdata,
title={On data visualization and interior design},
author={Brehmer, Matthew},
booktitle={Proc. IEEE VIS alt.VIS Workshop},
year={2025},
note={\href{https://altvis.github.io/papers/2025/abstracts/interior-design.pdf}{altvis.github.io/papers/2025/abstracts/interior-design.pdf}}
}

@misc{Graham2008,
title = {The geologic time spiral: A path to the past},
author = {Graham, Joseph and Newman, William and Stacy, John},
note = {\href{http://pubs.usgs.gov/gip/2008/58/}{pubs.usgs.gov/gip/2008/58}},
year = {2008}
}

@book{odell2024saving,
title={Saving Time: Discovering a Life Beyond the Clock},
author={Odell, Jenny},
year={2023},
publisher={Random House}
}

\end{document}